\documentclass[conference]{IEEEtran}
\usepackage{amsmath, amssymb}
\usepackage{graphicx}
\usepackage{cite}

\begin{document}

\title{Optimal Control for Network of Coupled Oscillators}

\author{Adnan Tahirovic} 

\maketitle

\begin{abstract}
This paper presents a nonlinear control framework for steering networks of coupled oscillators toward desired phase-locked configurations. Inspired by brain dynamics, where structured phase differences support cognitive functions, the focus is on achieving synchronization patterns beyond global coherence. The Kuramoto model, expressed in phase-difference coordinates, is used to describe the system dynamics. The control problem is formulated within the State-Dependent Riccati Equation (SDRE) framework, enabling the design of feedback laws through state-dependent factorisation. The unconstrained control formulation serves as a principled starting point for developing more general approaches that incorporate coupling constraints and actuation limits. Numerical simulations demonstrate that the proposed approach achieves robust phase-locking in both heterogeneous and large-scale oscillator networks, highlighting its potential applications in neuroscience, robotics, and distributed systems.
\end{abstract}

\section{Introduction}

The coordinated behavior of coupled oscillators is central to the operation of many natural and engineered systems. Examples range from biological rhythms such as heartbeat and sleep cycles to engineered systems like electric power grids and neural circuits in the brain~\cite{dorfler2014synchronization,strogatz2000kuramoto}. A common model used to analyze these synchronization phenomena is the Kuramoto model, which describes how oscillator phases evolve over time through nonlinear sinusoidal coupling~\cite{kuramoto1975self}.

\begin{figure}[!t] \centering \includegraphics[width=\linewidth]{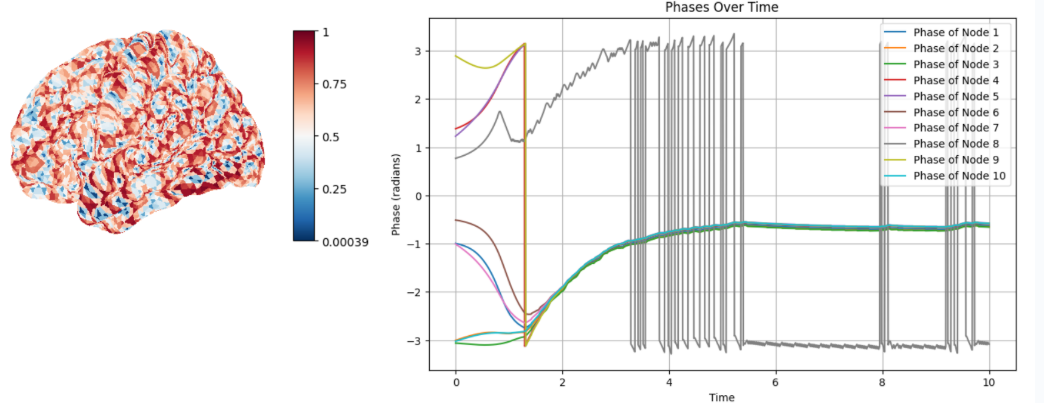} \caption{Illustrative example of the proposed optimal control driving a network of 10 oscillators to achieve a desired brain functional pattern by aligning all phases to prescribed values.} \label{fig:brain_fc_simulation} \end{figure}

In many cases, the objective is not simply to synchronize all oscillators to a common phase, but rather to steer the system toward a specific \textit{phase-locked configuration}, where certain phase differences are maintained. Such configurations can encode meaningful relationships among components. This is particularly relevant in neuroscience, where cognitive states are thought to emerge from structured patterns of phase coherence, referred to as \textit{functional connectivity}, between different regions of the brain, see. e.g. ~\cite{varela2001brain,lu2021framework}. For instance, distinct patterns of phase alignment may correspond to attention, memory, or resting-state brain activity. Reproducing these patterns in silico enables better understanding of brain dynamics and opens the door to targeted neuromodulation strategies. Fig.~\ref{fig:brain_fc_simulation} illustrates this concept: a control input is applied to a network of ten oscillators, each representing a brain region, to achieve a desired pattern of phase differences aligned with a functional target.

A similar need to control phase differences arises in power systems. There, phase differences between generators and loads determine power flow directions and magnitudes~\cite{dorfler2013synchronization}. Following disturbances or faults, guiding the system back to a specific phase-locked state ensures safe, stable, and efficient grid operation.

This work proposes a nonlinear optimal control framework for enforcing desired phase-locked states in networks of coupled oscillators. The Kuramoto model is used to describe the system dynamics, which is 
reformulated in terms of relative phase differences. This leads to a reduced-order representation that captures the essential behavior of the network. The proposed control strategy is based on the SDRE method~\cite{cloutier1997state}, which extends linear-quadratic control theory (LQR) to nonlinear systems by solving a Riccati equation at each state along the trajectory. 
The study focuses on the unconstrained control setting, offering a clear starting point for developing coupling-based strategies that can later accommodate actuation or physical limits. This choice reflects real-world scenarios where modulating coupling gains is feasible, such as in neuromorphic systems. At the same time, it lays the groundwork for future extensions to constrained or decentralized settings. Numerical simulations show that the proposed SDRE-based controller achieves reliable convergence to the target configuration across networks with varying sizes and diverse natural frequencies. 

The rest of the paper is structured as follows. Section~\ref{sec:model} introduces the Kuramoto model and derives the reduced-order error dynamics. Section~\ref{sec:control} presents the SDRE control framework and its adaptation to oscillator networks. Section~\ref{sec:results} demonstrates the controller’s performance through simulations, and Section~\ref{sec:conclusion} summarizes key findings and outlines directions for future work.

\section{Derivation of the Kuramoto Model and Error Dynamics}
\label{sec:model}
The dynamics of \( N \in \mathbb{Z}_{>1} \) coupled oscillators in the Kuramoto model, subject to multiplicative control, is given by:
\begin{equation}
\dot{\theta}_i = \omega_i + \frac{K}{N} u_i \sum_{j=1}^{N} \sin(\theta_j - \theta_i), \quad i = 1, \dots, N,
\end{equation}
where:
\begin{itemize}
    \item \( \theta_i \in \mathbb{R} \) is the phase of the \( i \)-th oscillator,
    \item \( \omega_i \in \mathbb{R} \) is its natural frequency,
    \item \( u_i \in \mathbb{R}_{>0} \) is a multiplicative control input,
    \item \( K \in \mathbb{R}_{>0} \) is the global coupling gain.
\end{itemize}
We define the multiplicative control input to adjust coupling strength as:
\begin{equation}
u_i = 1 + v_i,
\end{equation}
where \( v_i \in \mathbb{R} \) is the control deviation from the nominal value and \( \boldsymbol{v} = [v_1, \dots, v_N]^\top \in \mathbb{R}^N \) denotes the full control vector. 

To eliminate redundancy due to the model's invariance under global phase shifts, we define a minimal set of phase differences:
\begin{equation}
X_i = \theta_{i+1} - \theta_i, \quad i = 1, \dots, N-1,
\end{equation}
resulting in the reduced state vector \( \boldsymbol{X} = [X_1, \dots, X_{N-1}]^\top \in \mathbb{R}^{N-1} \). The number of independent differences is \( N - 1 \), since the sum of adjacent differences satisfies:
\[
\sum_{i=1}^{N-1} X_i = \theta_N - \theta_1,
\]
and the model remains unchanged under a global shift \( \theta_i \mapsto \theta_i + \alpha \), meaning that one phase can be fixed without loss of generality (e.g., $\theta_1=0$).

To track a desired phase configuration \( \boldsymbol{X}^{\text{des}} \in \mathbb{R}^{N-1} \), we define the error vector:
\begin{equation}
e_i = X_i - X_i^{\text{des}}, \quad \text{or} \quad \boldsymbol{e} = \boldsymbol{X} - \boldsymbol{X}^{\text{des}} \in \mathbb{R}^{N-1}.
\end{equation}
Differentiating the error yields:
\begin{equation}
\begin{aligned}
\dot{e}_i &= \dot{\theta}_{i+1} - \dot{\theta}_i \\
&= (\omega_{i+1} - \omega_i) + \frac{K}{N} \Bigg[ (1 + v_{i+1}) \sum_{j=1}^{N} \sin(\theta_j - \theta_{i+1}) \\
&\quad - (1 + v_i) \sum_{j=1}^{N} \sin(\theta_j - \theta_i) \Bigg].
\end{aligned}
\end{equation}
By defining the dynamics in vector form, we obtain:
\begin{equation}
\dot{\boldsymbol{e}} = \boldsymbol{f}(\boldsymbol{e}) + \boldsymbol{c} + B(\boldsymbol{e}) \boldsymbol{v},
\end{equation}
where:
\begin{itemize}
    \item \( \boldsymbol{c} \in \mathbb{R}^{N-1} \) is the vector of natural frequency differences:
    \[
    \boldsymbol{c} = 
    \begin{bmatrix}
    \omega_2 - \omega_1 \\
    \omega_3 - \omega_2 \\
    \vdots \\
    \omega_N - \omega_{N-1}
    \end{bmatrix},
    \]
    \item \( \boldsymbol{f}(\boldsymbol{e}) \in \mathbb{R}^{N-1} \) contains the nonlinear terms:
    \[
    f_i = \frac{K}{N} \left[ \sum_{j=1}^{N} \sin(\theta_j - \theta_{i+1}) - \sum_{j=1}^{N} \sin(\theta_j - \theta_i) \right],
    \]
    with \( \theta_i \) reconstructed from \( \boldsymbol{e} \) and \( \boldsymbol{X}^{\text{des}} \) using:
    \[
    \theta_i = \theta_1 + \sum_{k=1}^{i-1} (e_k + X_k^{\text{des}}), \quad \theta_1 = 0,
    \]
    \item \( B(\boldsymbol{e}) \in \mathbb{R}^{(N-1) \times N} \) is the control influence matrix defined elementwise as:
    \[
    B_{ij}(\boldsymbol{e}) =
    \begin{cases}
    -\frac{K}{N}\sum_{k=1}^{N} \sin(\theta_k - \theta_i), & j = i, \\
    +\frac{K}{N}\sum_{k=1}^{N} \sin(\theta_k - \theta_{i+1}), & j = i+1, \\
    0, & \text{otherwise}.
    \end{cases}
    \]
\end{itemize}

The full SDRE-based control implementation for guiding coupled Kuramoto oscillators is presented in the following section.

\section{Proposed SDRE-Based Control for Targeting Phase-Locked States in Coupled Kuramoto Oscillators}
\label{sec:control}

\subsection{SDRE-Based Control for Continuous-time Systems}
The SDRE method offers a structured and computationally tractable approach for designing feedback controllers for nonlinear systems (see, e.g., \cite{c1}). Among nonlinear control strategies, it stands out for its ability to generalize LQR principles while remaining suitable for real-time implementation (see, e.g. ~\cite{c16,c161}). The SDRE framework employs state-dependent coefficient (SDC) matrices to embed the nonlinear dynamics into a locally linear representation through appropriate system factorization, enabling the pointwise solution of Riccati equations. This structure preserves key nonlinear characteristics of the system, making SDRE especially effective for time-varying or highly nonlinear dynamics where traditional linearization fails to deliver satisfactory performance.

In addition to control design, the SDRE methodology has also been applied to state estimation in nonlinear systems through a unified SDRE-based control and estimation framework~\cite{tahirovic2025optimal}. This highlights the method’s broader relevance in both control and estimation settings. Furthermore, recent advances have integrated SDRE with policy iteration techniques to improve convergence and optimality in nonlinear control~\cite{c162,c163}, while robust extensions, including combinations with integral sliding mode control, have enhanced resilience under uncertainty~\cite{c164}.

The fundamental principle of SDRE-based control lies in transforming the nonlinear system dynamics into a linear-like representation through factorization using SDC matrices.
\begin{equation}
    \dot{x} = A(x)x + B(x)u
\end{equation}
where \(x \in \mathbb{R}^n\) is the state vector, \(u \in \mathbb{R}^m\) is the control input, \(A(x) \in \mathbb{R}^{n \times n}\) is the state-dependent system matrix, and \(B(x) \in \mathbb{R}^{n \times m}\) is the state-dependent input matrix.

This form enables the application of Riccati-based optimization to nonlinear systems by solving a modified version of the standard LQR problem. The performance objective typically involves minimizing an infinite-horizon cost function defined as:
\begin{equation}
    J = \frac{1}{2}\int_{0}^{\infty}\left(x^TQ(x)x + u^TR(x)u\right)dt,
\end{equation}
where \(Q(x) = D^T(x)D(x) \geq 0\) and \(R(x) > 0\) ensure positive semidefiniteness and positive definiteness, respectively.

The resulting feedback control law takes the form:
\begin{equation}
    u(x) = -K(x)x = -R^{-1}(x)B^T(x)P(x)x,
\end{equation}
where \(P(x)\) solves the following state-dependent algebraic Riccati equation:
\begin{align}
    &P(x)A(x) + A^T(x)P(x) \nonumber \\
    &\quad - P(x)B(x)R^{-1}(x)B^T(x)P(x) + Q(x) = 0
\end{align}
To ensure the feasibility of the solution, controllability of the pair \((A(x), B(x))\) must be verified. This is typically assessed through the rank condition of the controllability matrix:
\begin{equation}
\text{rank}\left[B(x) \ A(x)B(x) \ \cdots \ A(x)^{n-1}B(x)\right] = n,
\end{equation}
where \(n\) is the number of states in the system.

An important feature of the SDRE method is the flexibility in SDC matrix selection. For a given nonlinear vector field \(f(x)\), multiple factorizations may exist, satisfying \(f(x) = A_1(x)x = A_2(x)x\). This non-uniqueness enables convex combinations such as
\begin{equation}
    A(x,\alpha) = \alpha A_1(x) + (1 - \alpha)A_2(x)
\end{equation}
to be valid factorizations as well, providing additional tuning possibilities~\cite{c1}.

While SDRE resembles the LQR framework in its use of Riccati-based control design~\cite{c1}, it should be noted that it is not derived from the Hamilton–Jacobi–Bellman equation. As such, it does not guarantee global optimality for nonlinear systems~\cite{c162, c163}, but nonetheless provides an effective and intuitive control strategy in many practical applications.

\subsection{Application of SDRE Control to the Kuramoto Oscillator Model}
\label{sdre_kuramoto}

In this section, we propose a nonlinear optimal control framework based on the SDRE-based control to regulate Kuramoto oscillator networks. The goal is to design a coupled control strategy such that the system reaches a desired phase-locked configuration, characterized by target differences \( \boldsymbol{X}^{\text{des}} \in \mathbb{R}^{N-1} \).

Recall that the reduced-order dynamics of the error vector \( \boldsymbol{e}(t) \in \mathbb{R}^{N-1} \), defined as \( e_i = \theta_{i+1} - \theta_i - X_i^{\text{des}} \), is given by:
\[
\dot{\boldsymbol{e}} = \boldsymbol{f}(\boldsymbol{e}) + \boldsymbol{c} + B(\boldsymbol{e}) \boldsymbol{v},
\]
where:
\begin{itemize}
    \item \( \boldsymbol{f}(\boldsymbol{e}) \in \mathbb{R}^{N-1} \) represents the nonlinear coupling interactions,
    \item \( \boldsymbol{c} \in \mathbb{R}^{N-1} \) is a constant vector representing the natural frequency differences, with \( c_i = \omega_{i+1} - \omega_i \),
    \item \( B(\boldsymbol{e}) \in \mathbb{R}^{(N-1)\times N} \) is the control input matrix,
    \item \( \boldsymbol{v} \in \mathbb{R}^{N} \) denotes the deviation from the nominal unit control input, so that \( \boldsymbol{u} = \boldsymbol{1} + \boldsymbol{v} \).
\end{itemize}

To apply the SDRE framework, we approximately factorize the nonlinear term \( \boldsymbol{f}(\boldsymbol{e}) \) using a first-order Taylor expansion:
\[
\boldsymbol{f}(\boldsymbol{e}) \approx \boldsymbol{f}(\boldsymbol{0}) + A(\boldsymbol{e}) \boldsymbol{e},
\]
where \( A(\boldsymbol{e}) = \frac{\partial \boldsymbol{f}}{\partial \boldsymbol{e}} \in \mathbb{R}^{(N-1)\times(N-1)} \) is the Jacobian matrix evaluated at each time step. This approximation allows us to cast the dynamics into the SDRE-compatible form:
\[
\dot{\boldsymbol{e}} \approx A(\boldsymbol{e}) \boldsymbol{e} + B(\boldsymbol{e}) \boldsymbol{v} + \underbrace{\left( \boldsymbol{f}(\boldsymbol{0}) + \boldsymbol{c} \right)}_{\text{bias}}.
\]
Since the term \( \boldsymbol{f}(\boldsymbol{0}) + \boldsymbol{c} \) introduces a constant bias, we define a compensating control offset:
\[
\boldsymbol{v}_{\text{bias}} = - B^\dagger(\boldsymbol{e}) \left[ \boldsymbol{f}(\boldsymbol{0}) + \boldsymbol{c} \right],
\]
where \( B^\dagger(\boldsymbol{e}) \) is the Moore–Penrose pseudoinverse of \( B(\boldsymbol{e}) \). The final control input is thus:
\[
\boldsymbol{v} = \boldsymbol{v}_{\text{bias}} + v_{SDRE},
\]
where \(v_{SDRE}\) is the SDRE-related control term.

The objective is to minimize the infinite-horizon quadratic cost functional:
\begin{equation}
J = \int_0^\infty \left[ \boldsymbol{e}(t)^\top Q \boldsymbol{e}(t) + \boldsymbol{v}(t)^\top R \boldsymbol{v}(t) \right] dt.
\end{equation}
Solving the continuous-time algebraic Riccati equation:
\begin{equation}
A^\top(\boldsymbol{e}) P + P A(\boldsymbol{e}) - P B(\boldsymbol{e}) R^{-1} B^\top(\boldsymbol{e}) P + Q = 0
\end{equation}
yields the positive-definite matrix \( P(\boldsymbol{e}) \) used to compute 
\begin{equation}
    v_{SDRE}(e) = -K(e)e = -R^{-1}(x)B^T(e)P(e)e.
\end{equation}

The Jacobian matrix \( A(\boldsymbol{e}) \) is constructed by first reconstructing the oscillator phases from the reduced coordinates using \( \theta_1 = 0 \) as a reference and computing
\[
\theta_k = \sum_{l=1}^{k-1} (e_l + X_l^{\text{des}}), \quad k = 2, \dots, N.
\]
Each entry \( A_{ij}(\boldsymbol{e}) \) is then obtained by differentiating the full nonlinear error dynamics with respect to \( e_j \), leading to:
\begin{align}
A_{ij}(\boldsymbol{e}) &= \frac{K}{N} \sum_{k=1}^{N} \Big[
\cos(\theta_k - \theta_{i+1}) \cdot \frac{\partial (\theta_k - \theta_{i+1})}{\partial e_j}
\nonumber \\
&\quad - \cos(\theta_k - \theta_i) \cdot \frac{\partial (\theta_k - \theta_i)}{\partial e_j}
\Big].
\end{align}

\section{Results}
\label{sec:results}
\subsection{Illustrative Example of Targeting Phase-Locked States}

We begin by evaluating the ability of the SDRE-based control framework to guide a network of coupled oscillators toward a predefined phase-locked configuration. In this example, we consider a system of \( N = 4 \) oscillators, each governed by the Kuramoto dynamics with multiplicative control and non-zero natural frequencies. The natural frequencies were fixed as \( \boldsymbol{\omega} =[1.30,\,1.39,\,0.44,\,1.28]^\top \), and the desired phase differences were chosen as \( \boldsymbol{X}^{\text{des}} =[-0.74,\,0.27,\,0.15]^\top \). The initial oscillator phases were set to \( \boldsymbol{\theta}(0) = [ 0.60,\,0.86,\,0.84,\,-0.13]^\top \).

The SDRE controller was configured with coupling strength \( K = 1.0 \), a strong penalty on phase error via \( Q = 1000 \cdot I_{3} \), and a moderate penalty on control effort via \( R = I_4 \), with \( I_N \) denoting the identity matrix of size \(N\)x\(N\). The simulation was run over a time horizon of \( T = 2 \) seconds with a step size of \( \Delta t = 0.01 \).

\begin{figure}[!ht]
    \centering
    \includegraphics[width=\linewidth]{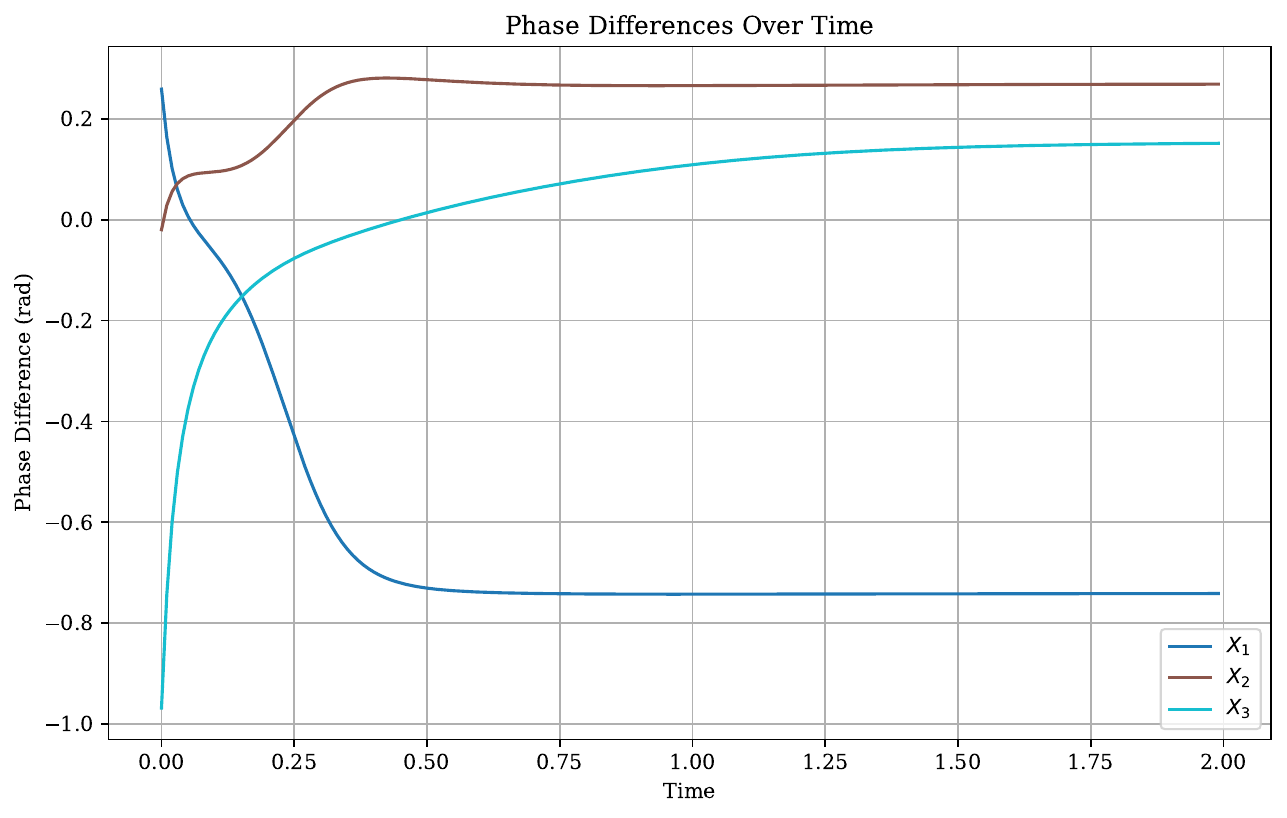}
    \caption{Phase differences \( X_i(t) \) for \( N = 4 \). All values converge to the corresponding targets in \( \boldsymbol{X}^{\text{des}} =[-0.74,\,0.27,\,0.15]^\top \).}
    \label{fig:phase_diff}
\end{figure}

\begin{figure}[!ht]
    \centering
    \includegraphics[width=\linewidth]{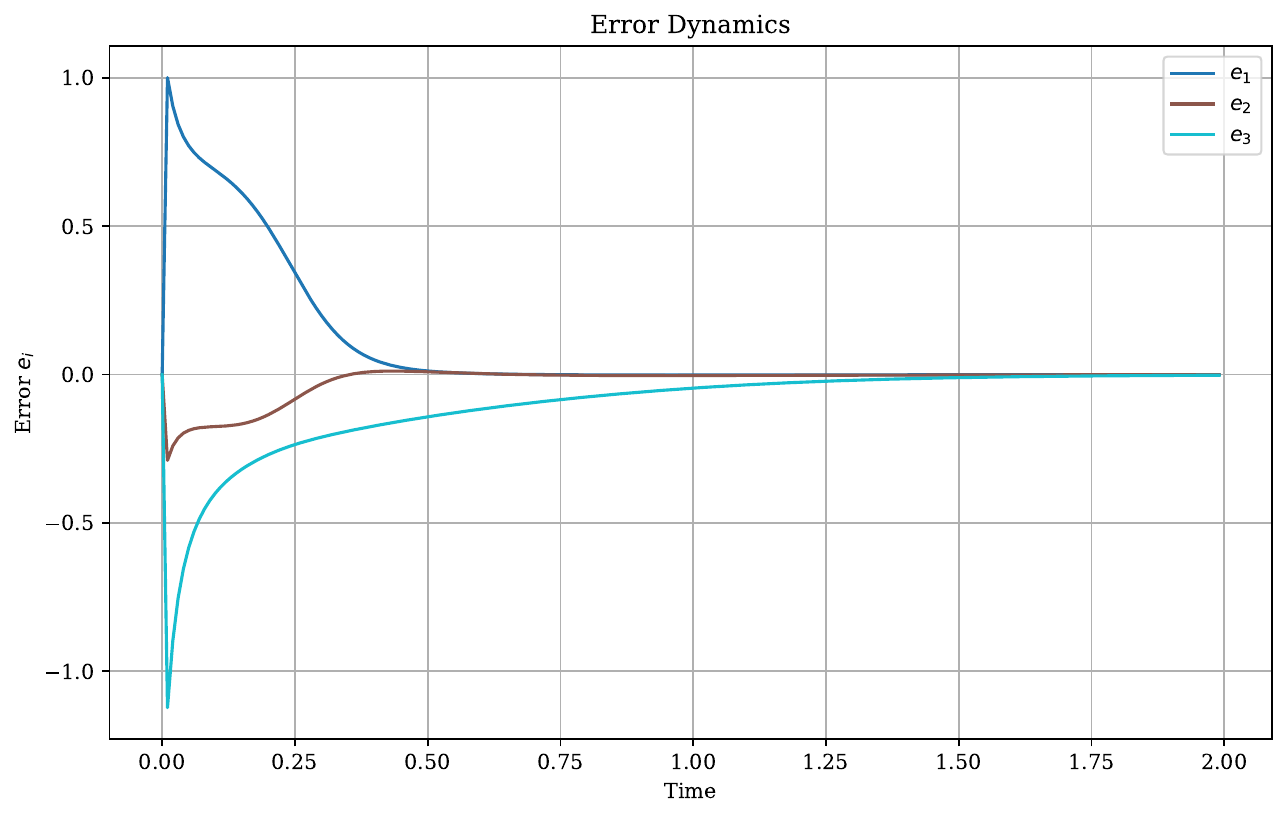}
    \caption{Error dynamics \( e_i(t) = X_i(t) - X_i^{\text{des}} \) for \( N = 4 \) showing smooth convergence to zero.}
    \label{fig:error_dynamics}
\end{figure}

Fig.~\ref{fig:phase_diff} shows the evolution of the minimal set of phase differences \( X_i = \theta_{i+1} - \theta_i \). All curves converge to the specified values in \( \boldsymbol{X}^{\text{des}} \), confirming that the SDRE controller is able to enforce the desired relative phase configuration. The error signals \( e_i(t) = X_i(t) - X_i^{\text{des}} \), shown in Fig.~\ref{fig:error_dynamics}, decay smoothly to zero, indicating successful synchronization.

\begin{figure}[!ht]
    \centering
    \includegraphics[width=\linewidth]{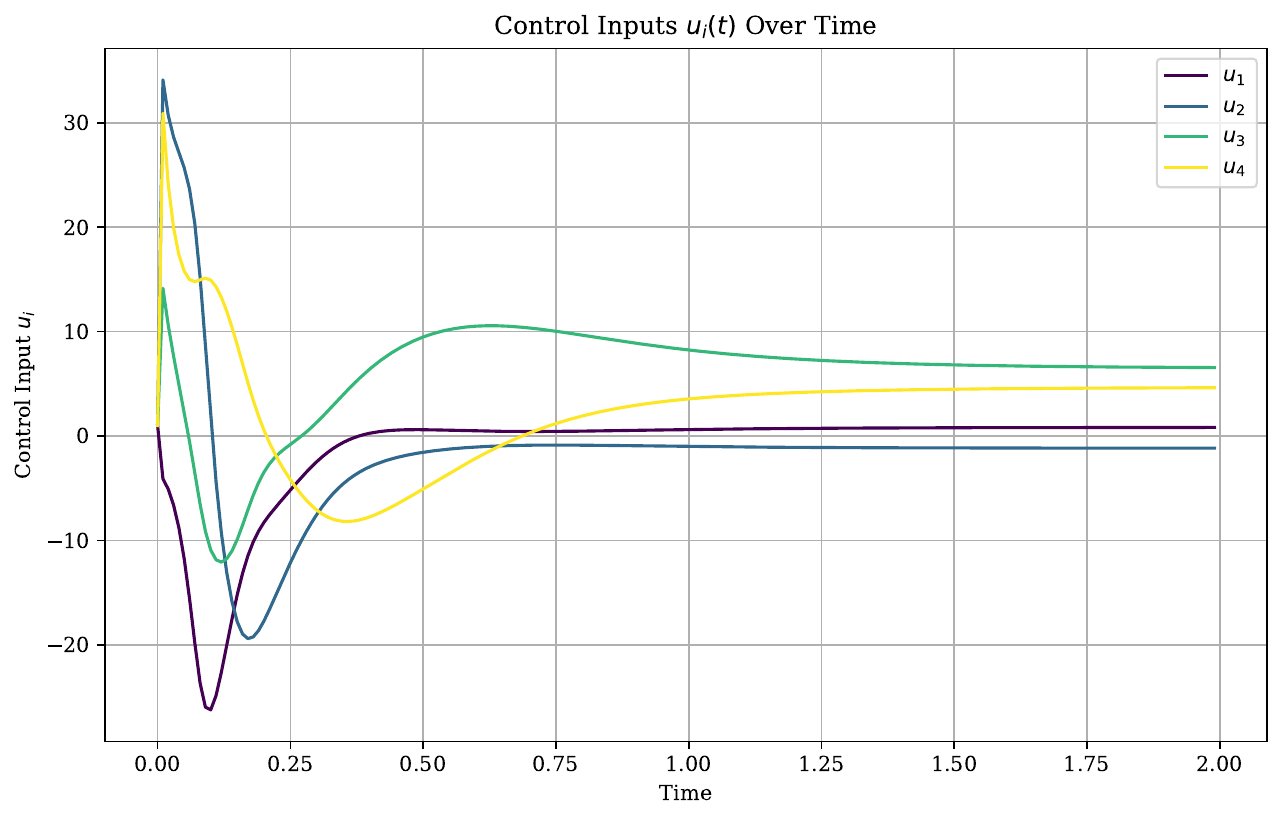}
    \caption{Control inputs \( u_i(t) \) for \( N = 4 \) remain smooth and bounded over time, \(\lim_{t \to \infty} \boldsymbol{u}(t) = [0.82,\,-1.16,\,6.56,\,4.63]^\top\).}
    \label{fig:control_inputs}
\end{figure}

The control signals  (Fig.~\ref{fig:control_inputs}) converges to steady values \(\lim_{t \to \infty} \boldsymbol{u}(t) = [0.82,\,-1.16,\,6.56,\,4.63]^\top\), indicating the coupling strengths required between oscillators to achieve the desired phase-locked state.

\subsection{Control Performance Under Extreme Frequency Dispersion}

To evaluate the limits of the proposed SDRE control framework, we simulate a case with highly dispersed natural frequencies:
\[
\omega = [0,\ \pi/3,\ 2\pi/3,\ \pi]^\top.
\]
The simulation is carried out for \( N = 4 \) oscillators with coupling strength \( K = 1.0 \). The desired phase differences are set to \( X^{\text{des}} = [-0.7,\ 1.2,\ -0.5]^\top \), and the initial phases are \( \theta(0) = [2.75,\ -0.96,\ 1.97,\ 2.10]^\top \). The simulation runs for \( T = 2 \) seconds with a time step of \( \Delta t = 0.01 \). The SDRE cost function uses weighting matrices \( Q = 1000 \cdot I_3 \) and \( R = I_4 \).

\begin{figure}[ht]
    \centering
    \includegraphics[width=\linewidth]{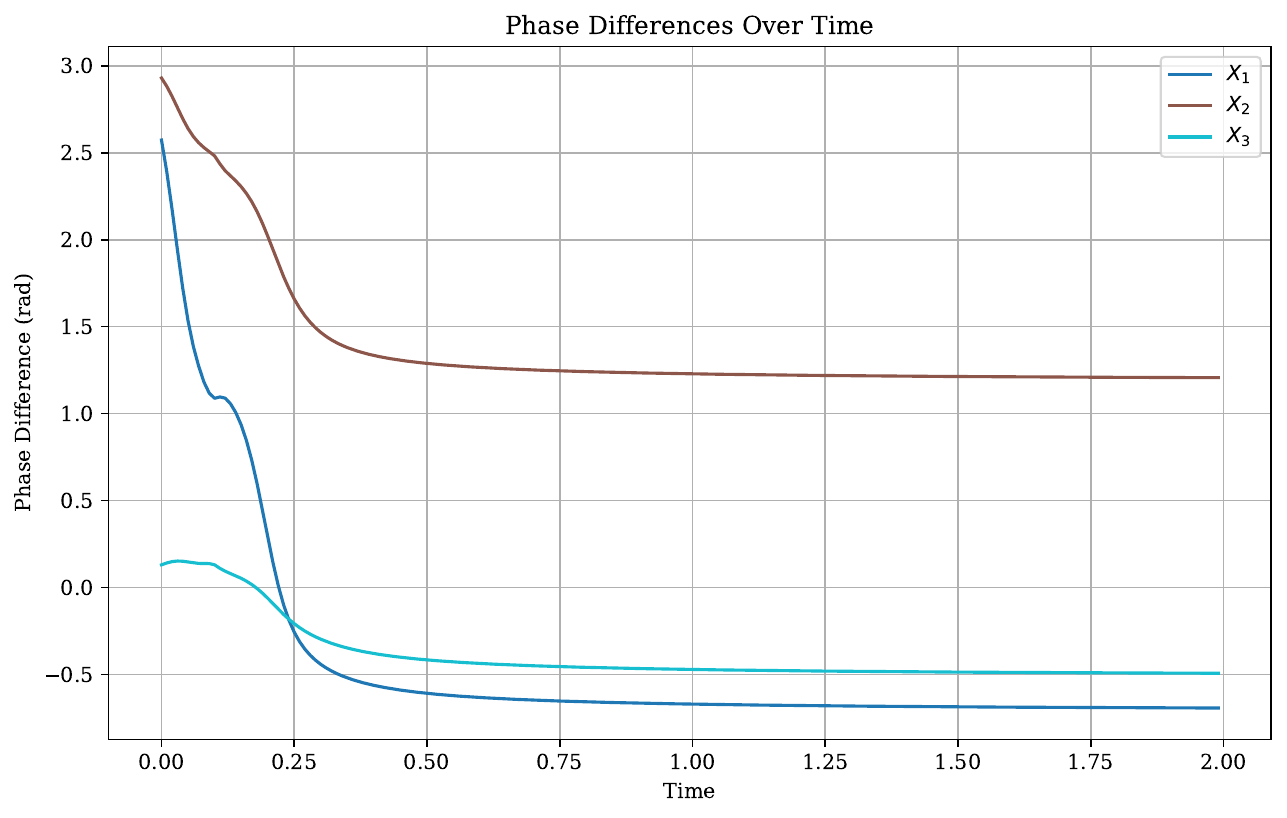}
    \caption{Evolution of phase differences \( X_i(t) \) for \( N = 4 \) under high frequency dispersion. Each curve converges to its respective desired value \( X^{\text{des}} = [-0.7,\ 1.2,\ -0.5]^\top \).}
    \label{fig:phase_diff_dispersion}
\end{figure}

\begin{figure}[ht]
    \centering
    \includegraphics[width=\linewidth]{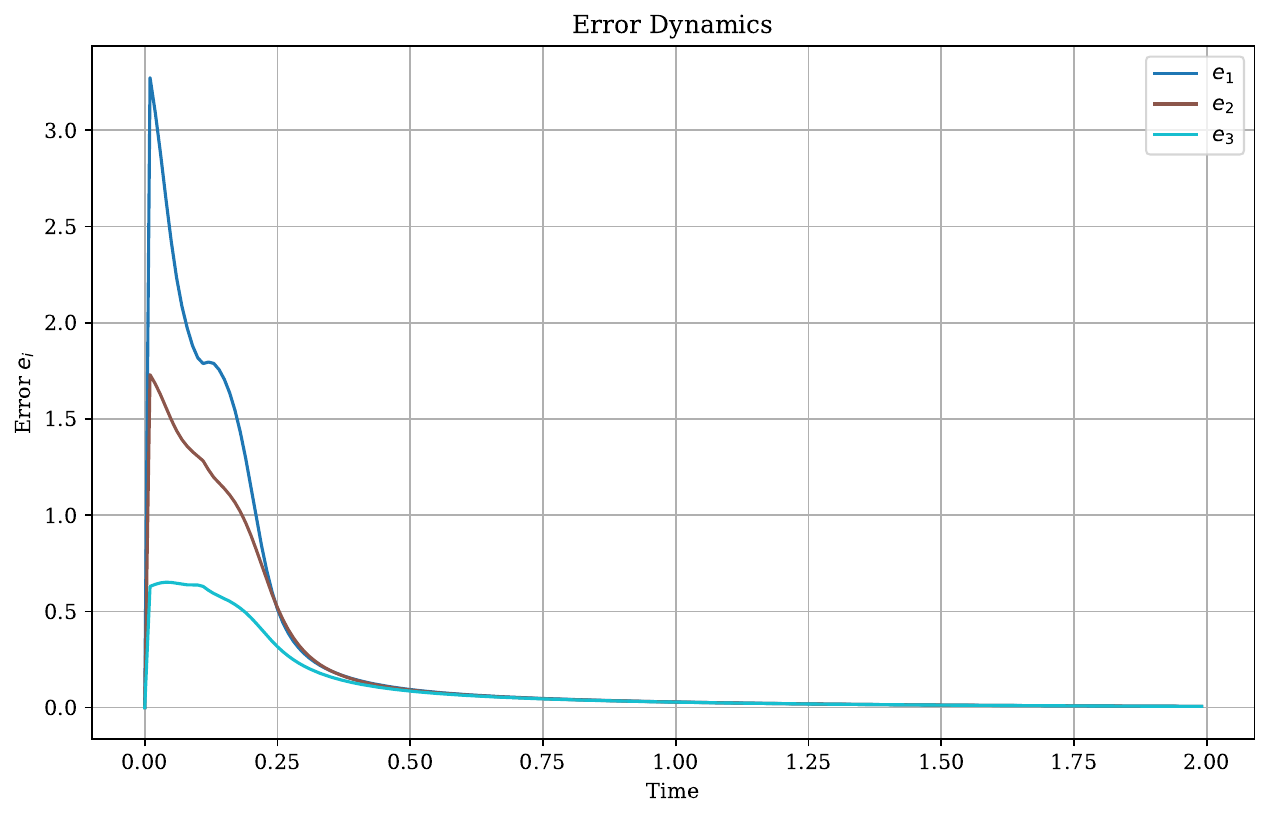}
    \caption{Error dynamics \( e_i(t) \) for \( N = 4 \) corresponding to the deviation from desired phase locking. Errors decay to zero despite frequency mismatch.}
    \label{fig:error_dispersion}
\end{figure}

\begin{figure}[ht]
    \centering
    \includegraphics[width=\linewidth]{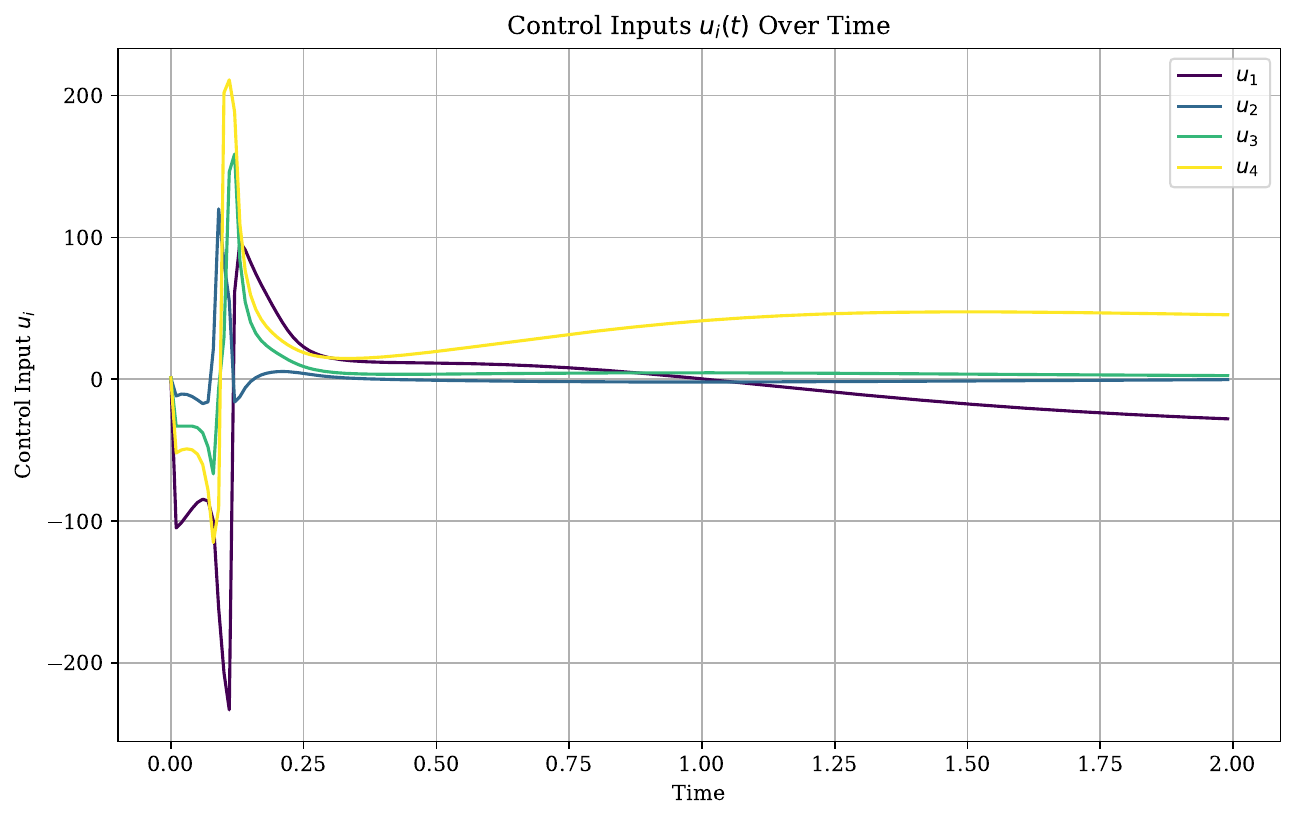}
    \caption{Control inputs \( u_i(t) \) for \( N = 4 \) over time. The controller compensates for natural frequency heterogeneity through time-varying actuation and converges to \(\lim_{t \to \infty} \boldsymbol{u}(t) = [-27.90,\,-0.28,\,2.55,\,45.44]^\top\).}
    \label{fig:control_u_dispersion}  
    \end{figure}

Despite the substantial mismatch, the SDRE controller succeeds in achieving convergence of the error dynamics \( e_i(t) \to 0 \) for all \( i \) (Fig.~\ref{fig:error_dispersion}), with control inputs \( u_i(t) \) adapting accordingly (Fig.~\ref{fig:control_u_dispersion}). The phase difference trajectories \( X_i(t) \) converge to their respective targets \( X_i^{\text{des}} \), and the oscillators settle into a phase-locked configuration (Fig.~\ref{fig:phase_diff_dispersion}).

This test illustrates the robustness of the proposed framework under significant variability in natural frequencies, conditions typical of heterogeneous robotic teams, oscillator-based computing platforms, or synthetic biological circuits, where exact matching of intrinsic dynamics is often infeasible.

\subsection{Influence of Weighting Matrices on Control Magnitude}

\begin{figure}[ht]
    \centering
    \includegraphics[width=\linewidth]{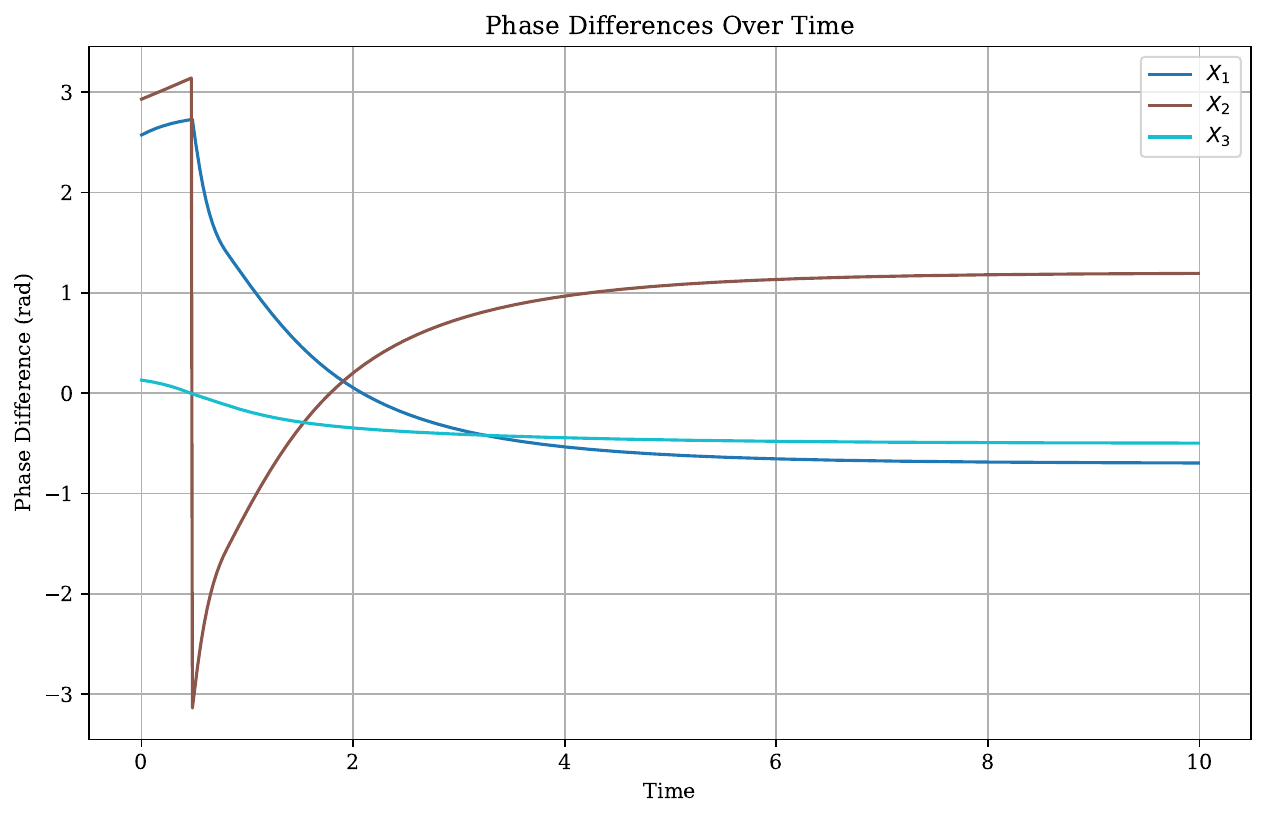}
    \caption{Phase differences \( X_i(t) \) for \( N = 4 \) under reduced error penalty (\( Q = 0.001 \cdot I_3 \)). Despite a lower synchronization priority, the trajectories remain well-behaved and approach the desired values, \( X^{\text{des}} = [-0.7,\ 1.2,\ -0.5]^\top \).}
    \label{fig:phase_diff_lowQ}
\end{figure}

\begin{figure}[ht]
    \centering
    \includegraphics[width=\linewidth]{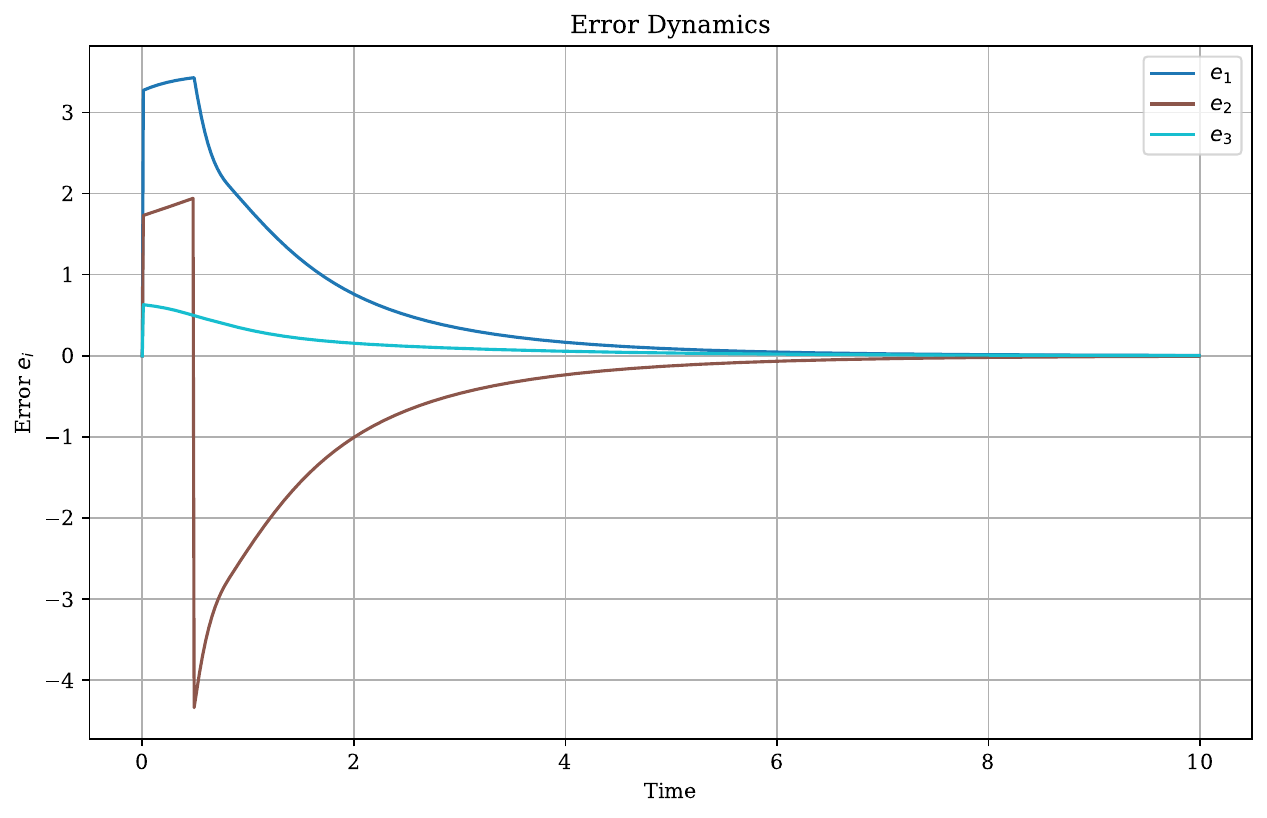}
    \caption{Error dynamics \( e_i(t) \) for \( N = 4 \) with low \( Q \). The convergence is slower but still achieves phase-locking within the simulation window.}
    \label{fig:error_lowQ}
\end{figure}

\begin{figure}[ht]
    \centering
    \includegraphics[width=\linewidth]{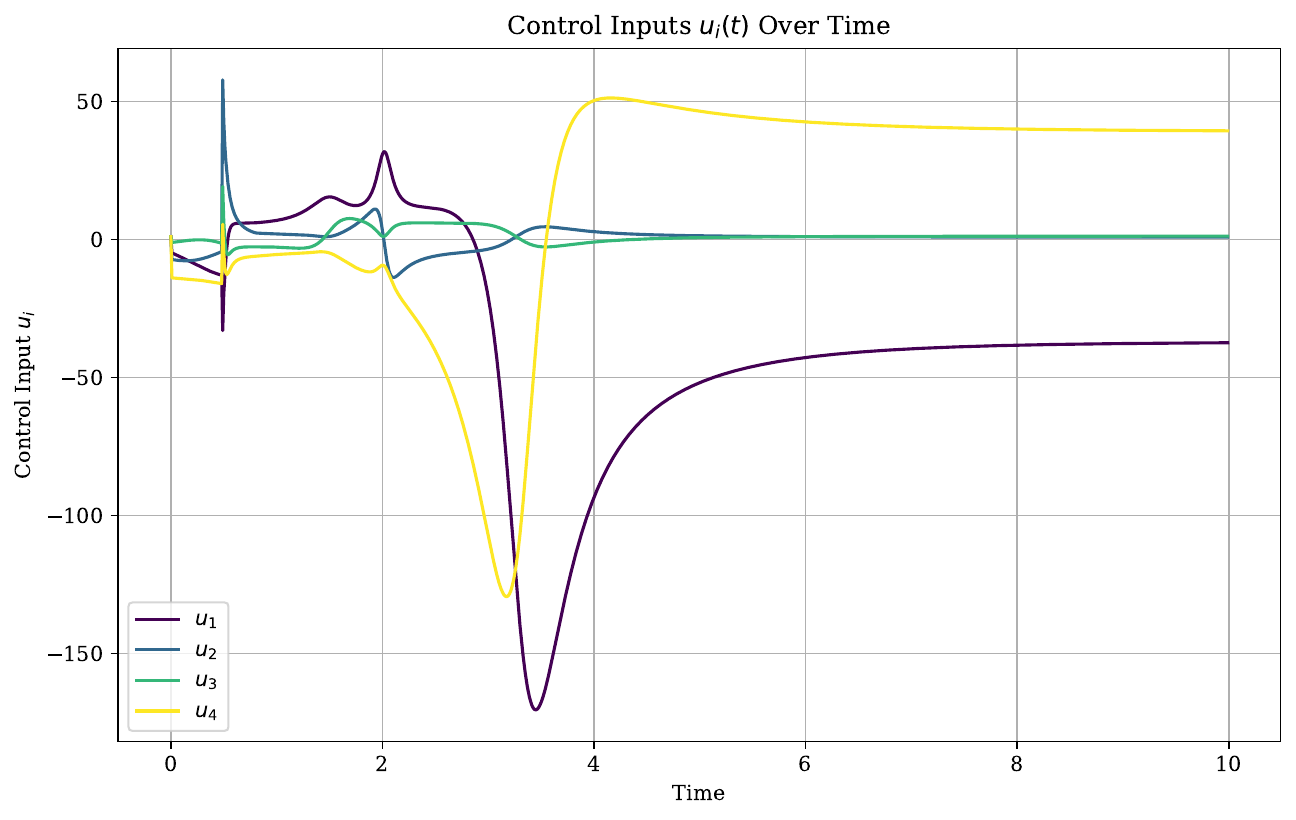}
    \caption{Control inputs \( u_i(t) \) for \( N = 4 \) under low \( Q \) weighting. Compared to the high-Q case, the control values remain significantly more bounded, illustrating the trade-off between control effort and tracking precision. \(\lim_{t \to \infty} \boldsymbol{u}(t) = [-37.43,\,0.87,\,1.19,\,39.36]^\top\).}
    \label{fig:control_u_lowQ}
\end{figure}

To illustrate how the relative weighting between synchronization accuracy and control effort influences the boundedness of control signals, an additional simulation is performed with a reduced error penalty matrix \( Q = 0.001 \cdot I_3 \), while keeping \( R = I_4 \), to compare it with the high-\( Q \) configuration from the previous subsection (Figs.~\ref{fig:phase_diff_lowQ} and~\ref{fig:error_lowQ}). This setting leads to significantly smaller control inputs \( \boldsymbol{u}(t) \), avoiding the large actuation magnitudes observed in the high-dispersion scenario (Fig.~\ref{fig:control_u_lowQ}). This outcome suggests that, to a certain extent, adjusting the ratio between \( Q \) and \( R \) in the SDRE framework can guide the controller toward producing more conservative control signals, thereby providing a limited form of implicit actuation regularization.

However, while such tuning improves boundedness, it does not guarantee feasibility across all settings. As a future direction, constrained control formulations of the SDRE method will be investigated to explicitly handle bounded control domains. This will enable reliable application of the framework to systems where physical, safety, or energy constraints place strict limits on actuation capabilities.

\subsection{Scalability to Larger Oscillator Networks}

To evaluate the scalability of the proposed SDRE-based control framework, Kuramoto oscillator networks of increasing size were simulated for \( N = 10 \), \( N = 20 \), \( N = 50 \), and \( N = 100 \). These experiments aim to assess whether the control law remains effective and computationally feasible as system size increases.

In all four scenarios, the coupling strength was fixed at \( K = 1.0 \), and natural frequencies \( \omega_i \) were sampled uniformly from the interval \( [0,\, \pi/2] \). Initial oscillator phases \( \theta_i(0) \) were drawn uniformly from the range \( [-\pi, \pi] \), and the desired phase differences \( X^{\text{des}}_i \in [-\pi/4, \pi/4] \) were randomly generated for each of the \( N-1 \) minimal state variables. Control weight matrices were selected as \( Q = 1000 \cdot I_{N-1} \) and \( R = I_N \), consistent with earlier simulation settings.

\begin{figure}[ht]
    \centering
    \includegraphics[width=\linewidth]{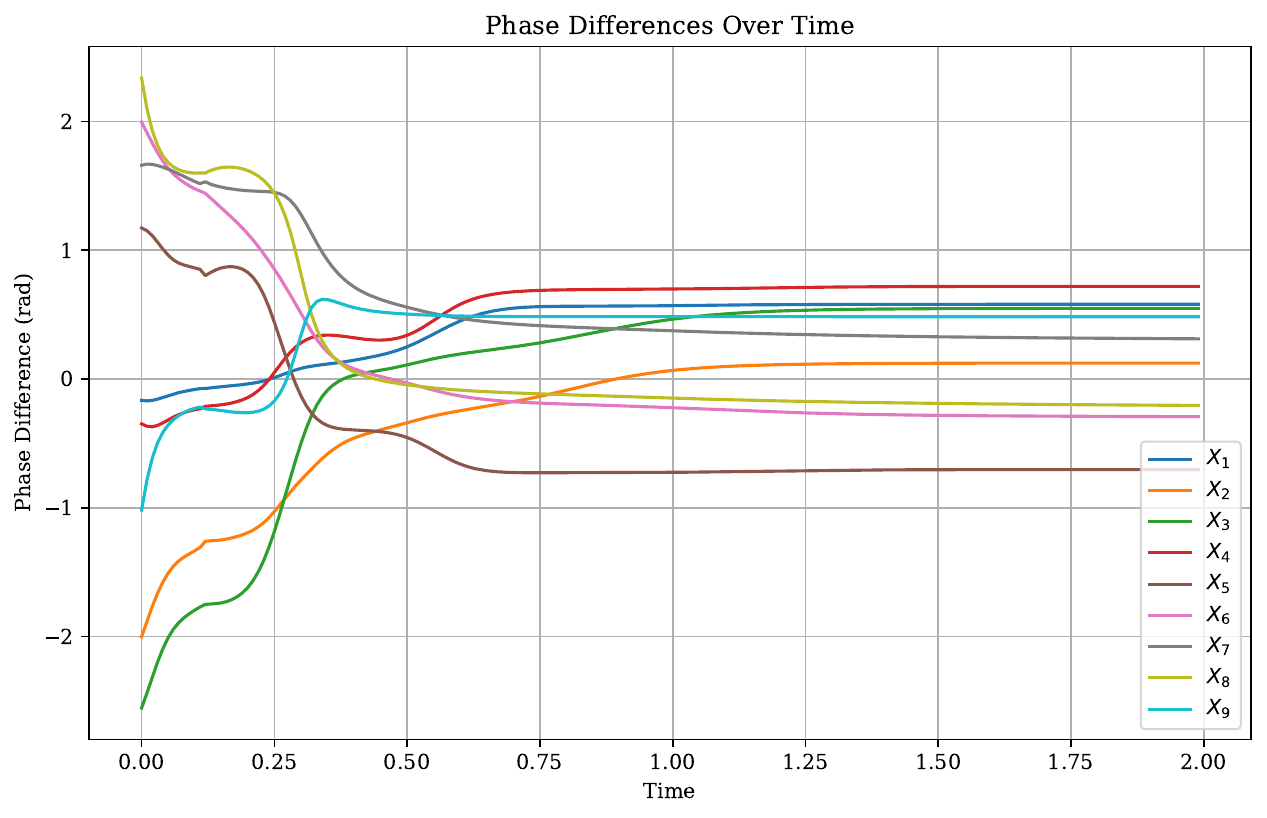}
    \caption{Phase differences \( X_i(t) \) for \( N = 10 \) oscillators.}
    \label{fig:phase_diff_N10}
\end{figure}

\begin{figure}[ht]
    \centering
    \includegraphics[width=\linewidth]{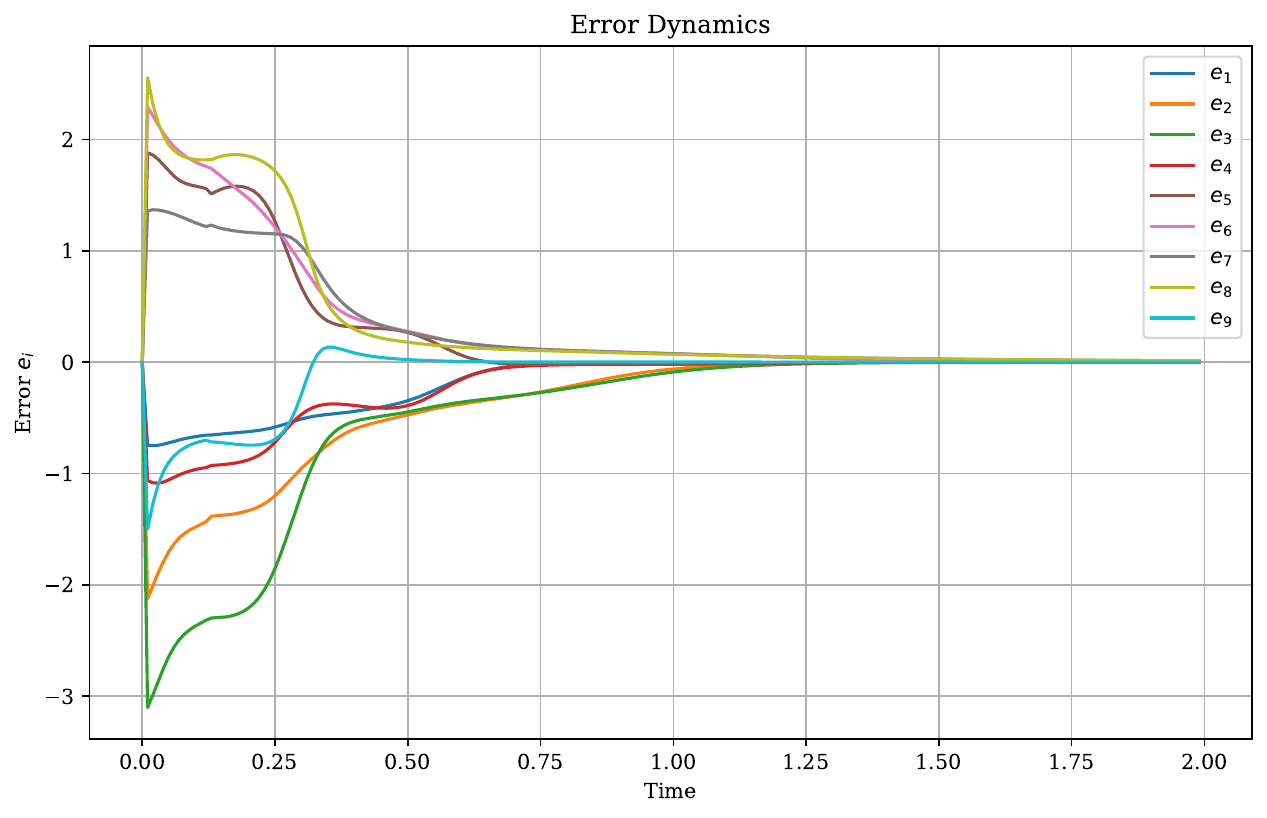}
    \caption{Error dynamics \( e_i(t) \) for \( N = 10 \).}
    \label{fig:error_N10}
\end{figure}

\begin{figure}[ht]
    \centering
    \includegraphics[width=\linewidth]{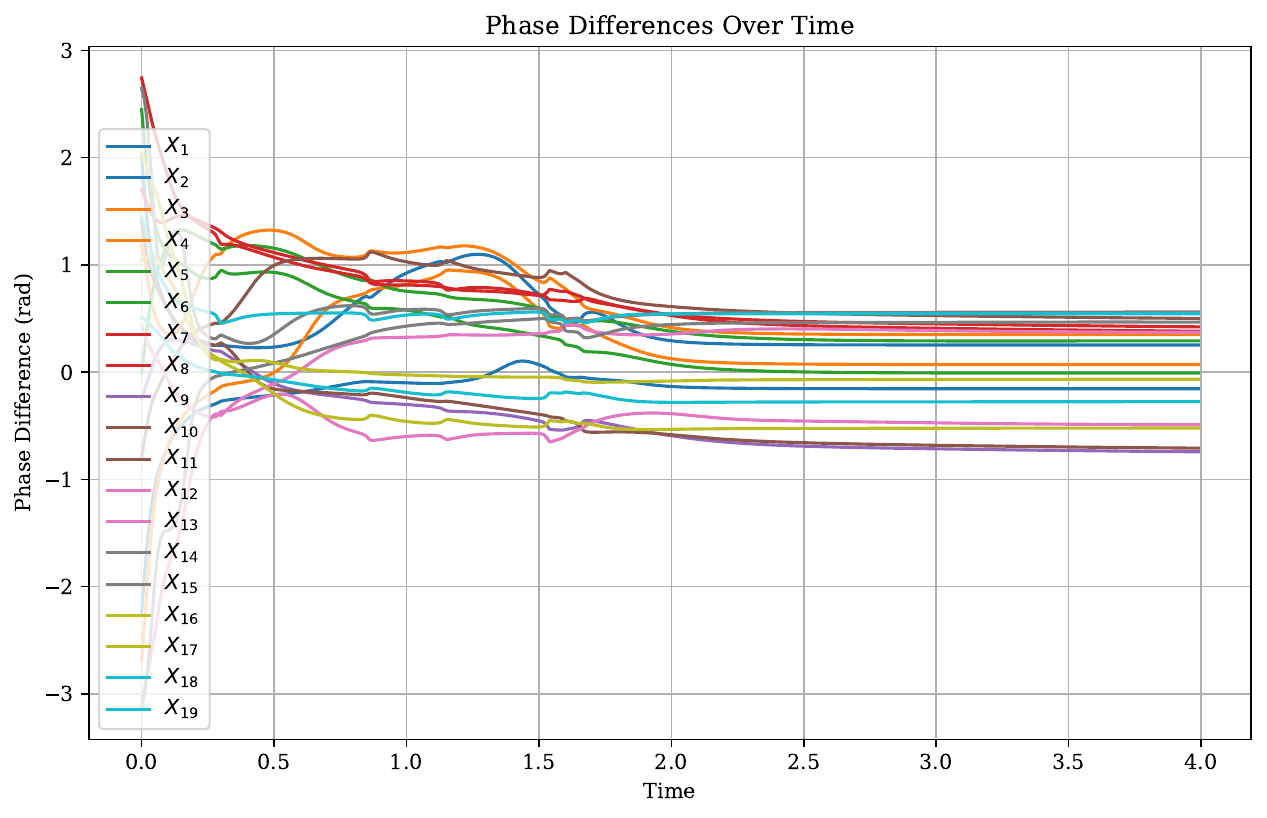}
    \caption{Phase differences \( X_i(t) \) for \( N = 20 \) oscillators.}
    \label{fig:phase_diff_N20}
\end{figure}

\begin{figure}[ht]
    \centering
    \includegraphics[width=\linewidth]{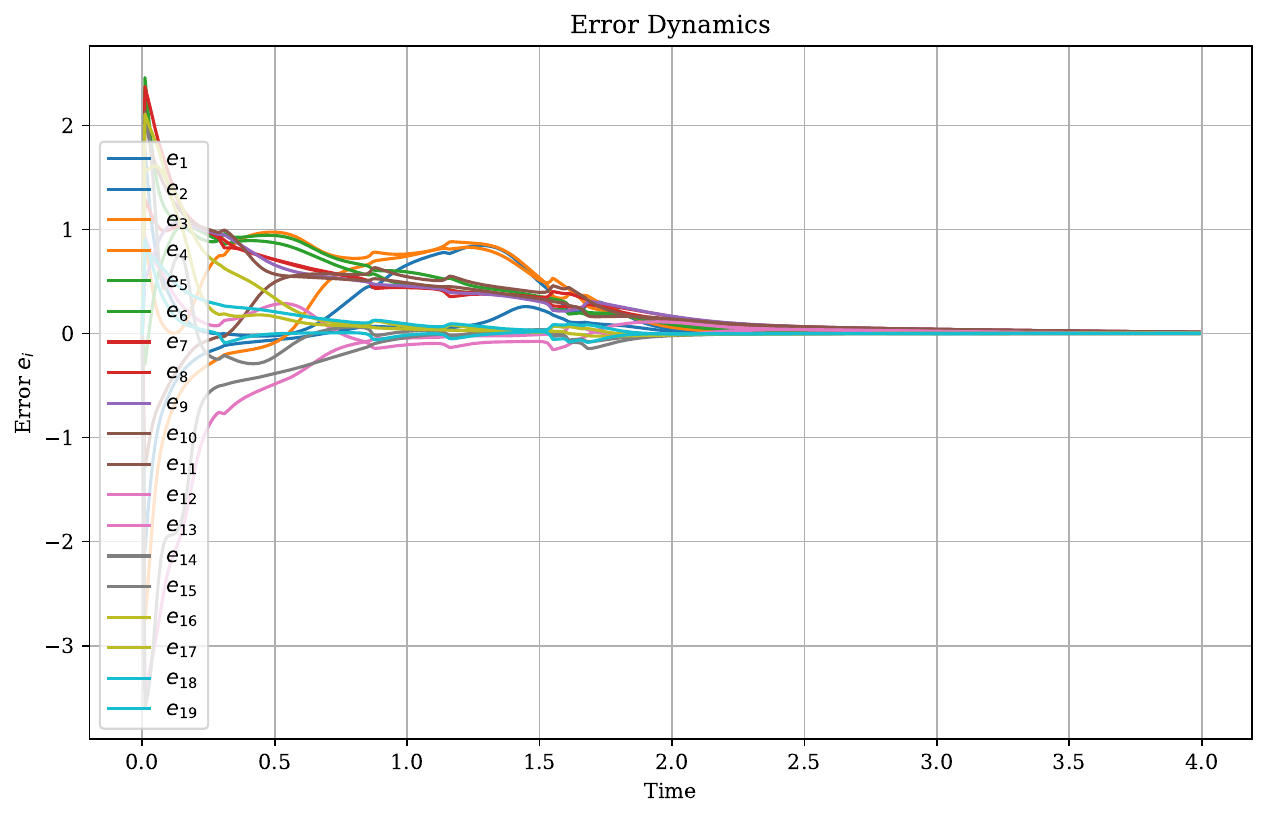}
    \caption{Error dynamics \( e_i(t) \) for \( N = 20 \).}
    \label{fig:error_N20}
\end{figure}

For \( N = 10 \) and \( N = 20 \), simulations confirm global convergence: phase differences \( X_i(t) \) successfully approach their respective targets \( X_i^{\text{des}} \), and the error dynamics \( e_i(t) \) decay to zero from arbitrary initial conditions. The final configurations of phase differences and error dynamics for these cases are illustrated in Figs.~\ref{fig:phase_diff_N10}–\ref{fig:error_N20}.

During initial testing, it was observed that for smaller \(N\), e.g. \( N \leq 20 \), the SDRE controller reliably achieved global convergence: the error dynamics \( \boldsymbol{e}(t) \) consistently decayed to zero across a broad range of initial conditions. However, for larger systems (\( N = 50 \) and \( N = 100 \)), simulations revealed that convergence was more sensitive to initialization. In particular, when the initial phase difference configuration was far from the target \( \boldsymbol{X}^{\text{des}} \), the system exhibited a persistent offset in the error dynamics. This behavior is attributed to the local nature of the first-order Taylor approximation used to factorize the nonlinear term \( \boldsymbol{f}(\boldsymbol{e}) \), which becomes less accurate over large deviations from equilibrium.

\begin{figure}[!t]
    \centering
    \includegraphics[width=\linewidth]{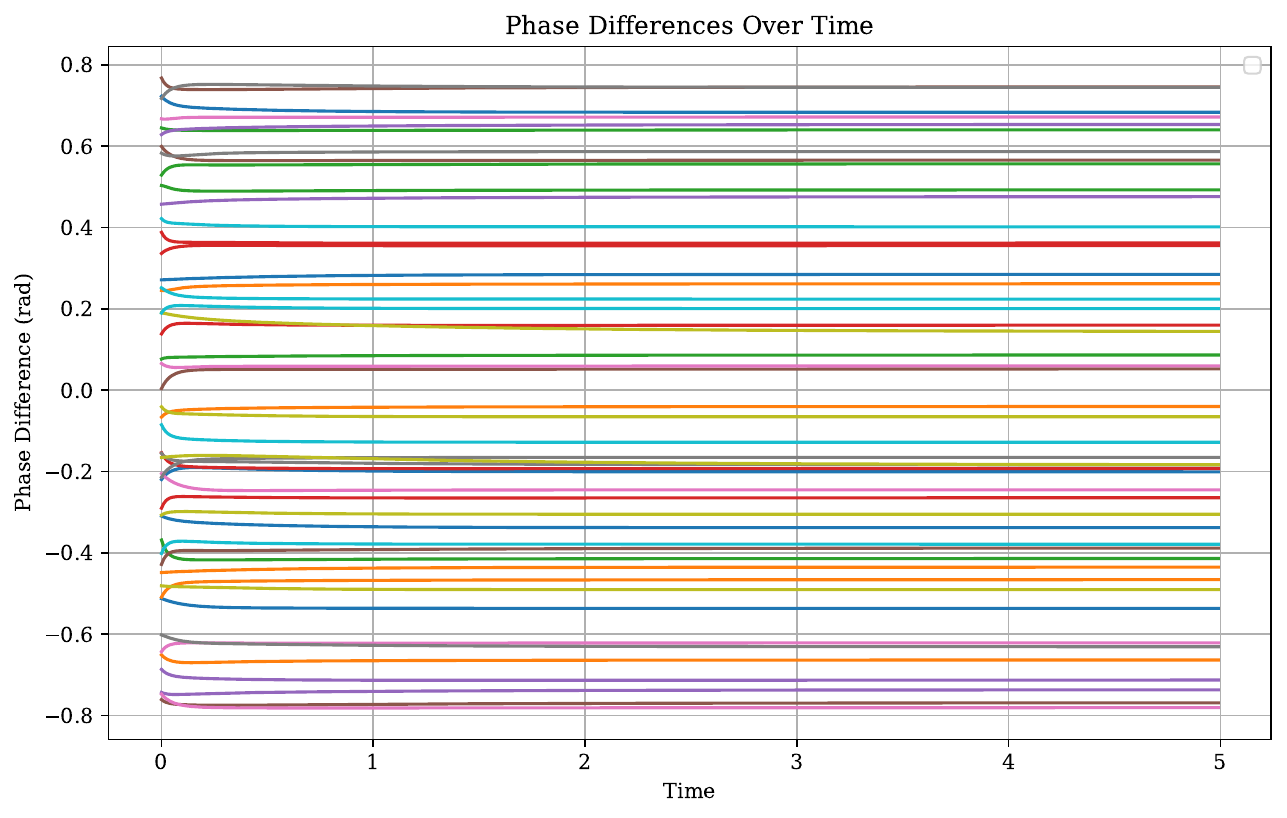}
    \caption{Phase differences \( X_i(t) \) for \( N = 50 \) oscillators.}
    \label{fig:phase_diff_N50}
\end{figure}

\begin{figure}[!t]
    \centering
    \includegraphics[width=\linewidth]{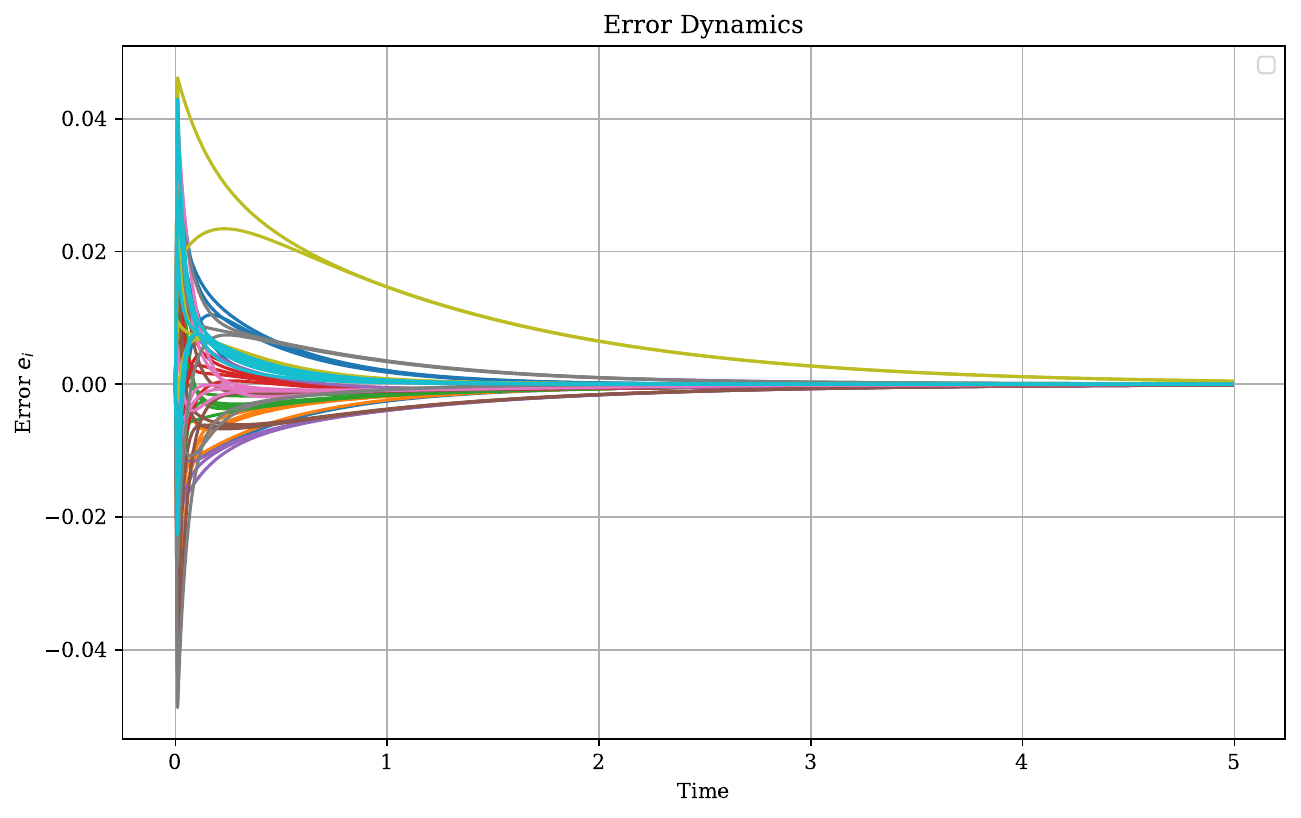}
    \caption{Error dynamics \( e_i(t) \) for \( N = 50 \).}
    \label{fig:error_N50}
\end{figure}

\begin{figure}[!t]
    \centering
    \includegraphics[width=\linewidth]{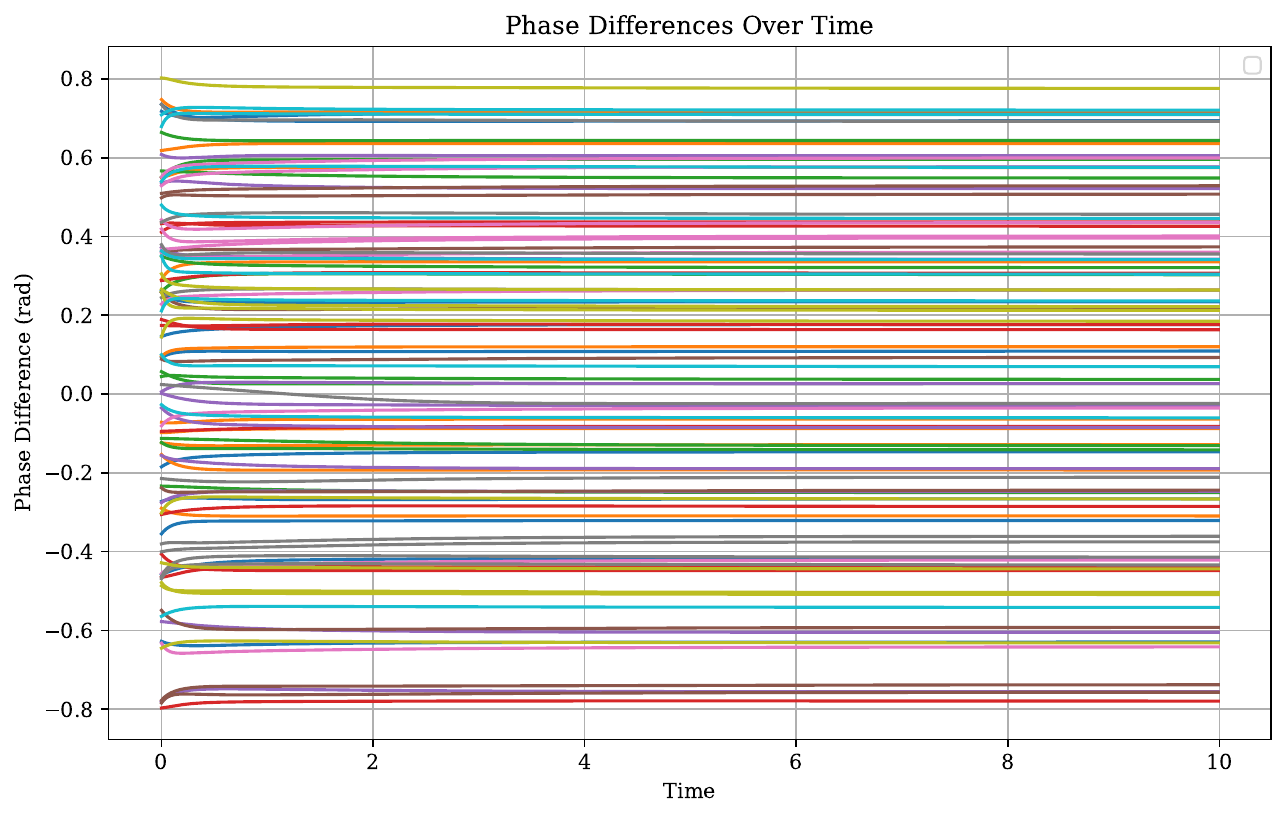}
    \caption{Phase differences \( X_i(t) \) for \( N = 100 \) oscillators.}
    \label{fig:phase_diff_N100}
\end{figure}

\begin{figure}[!t]
    \centering
    \includegraphics[width=\linewidth]{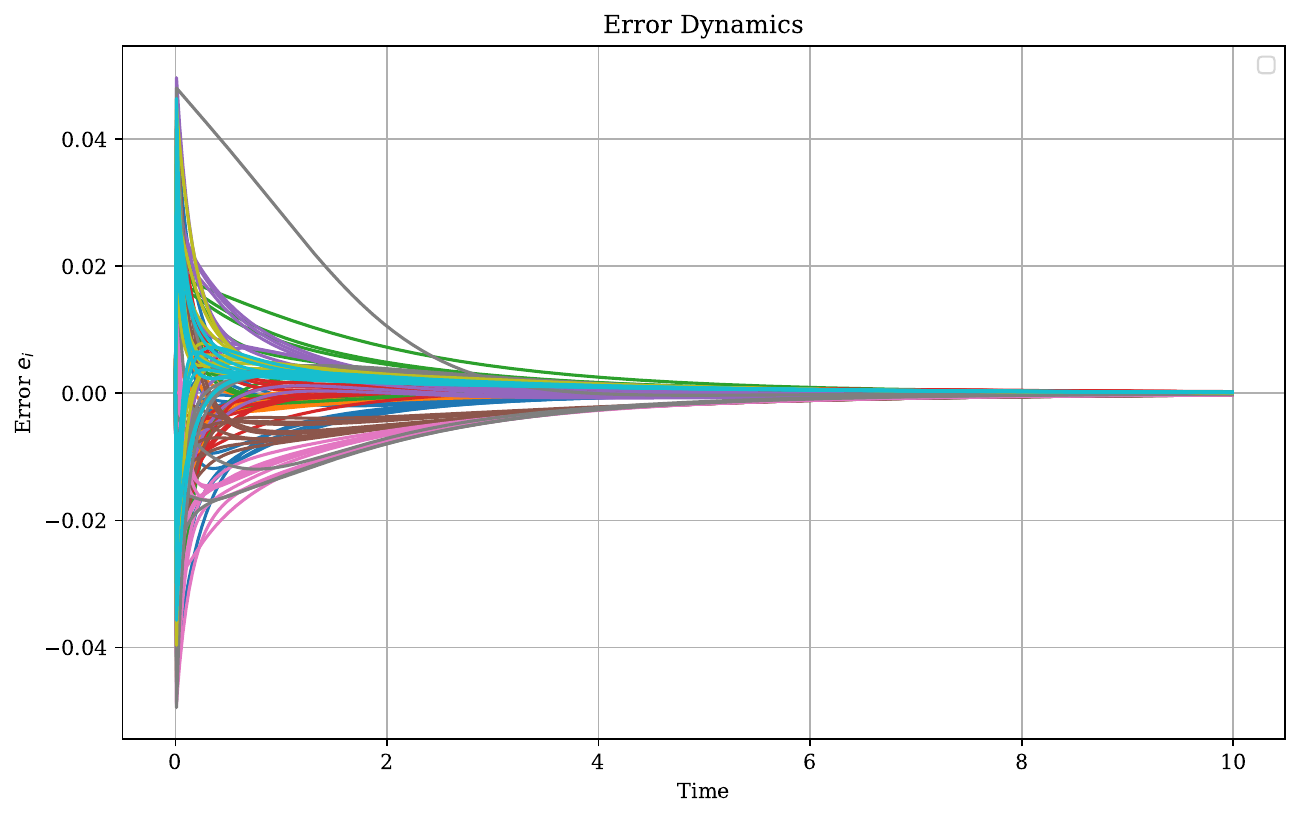}
    \caption{Error dynamics \( e_i(t) \) for \( N = 100 \).}
    \label{fig:error_N100}
\end{figure}

To mitigate this effect, simulations for \( N = 50 \) and \( N = 100 \) were initialized such that the phase difference states were close to \( \boldsymbol{X}^{\text{des}} \). Under these conditions, the error \( \boldsymbol{e}(t) \) successfully converged to zero, confirming that the SDRE framework can scale to large oscillator networks when the control task involves transitioning between phase-locked configurations that are relatively close. This reflects practical scenarios where the system must adapt between similar functional connectivity patterns, such as in brain dynamics. The results for these cases are shown in Figs.~\ref{fig:phase_diff_N50}–\ref{fig:error_N100}.

These results demonstrate that the SDRE framework maintains performance and numerical stability across increasing network sizes. While convergence is globally observed for small and medium networks, larger systems exhibit local convergence behavior, depending on the proximity of the initial state to the desired equilibrium.

\section{Conclusion} \label{sec:conclusion}

This paper introduced a nonlinear control strategy for guiding networks of coupled oscillators toward specified phase-locked configurations. By using the Kuramoto model and reformulating the system in terms of relative phase differences, the control problem was cast into a structure amenable to the SDRE framework. The nonlinear dynamics is approximately factorized through an exact Jacobian at each state, enabling the computation of time-varying feedback laws that respond to the current system behavior.

Simulation results confirmed that the proposed method reliably achieves the desired synchronization patterns, even in networks with heterogeneous natural frequencies and larger scales. The control inputs remained smooth and bounded, demonstrating the practical feasibility of the approach. The method’s scalability further supports its relevance for real-world applications.

This framework opens up possibilities for controlling functional dynamics in systems where phase relationships play a critical role, such as brain networks, oscillator-based computing, and coordinated robotics. As a next step, future work will focus on extending the approach to handle constraints on control inputs, making it more suitable for implementation in systems with physical or operational limitations.

\bibliographystyle{IEEEtran}
\bibliography{references}

\end{document}